\documentclass[12pt]{amsart}
\usepackage{amsmath,amssymb,amscd,latexsym,graphicx,epsfig}
\DeclareMathSizes{10}{10}{9}{9}

\newcommand{\ra}{\mathrm A}
\newcommand{\rb}{\mathrm B}

\newcommand{\uA}{\uparrow}
\newcommand{\dA}{\downarrow}

\title{THE FREE WILL THEOREM} 
\author{John Conway and Simon Kochen}
\address{Princeton University, Department of Mathematics, Princeton, NJ  08544-1000} 
\dedicatory{}
\email{conway@math.princeton.edu \\ kochen@math.princeton.edu}
\thanks{}
\date{March 31, 2006}

\begin{document}

\abstract

On the basis of three physical axioms, we prove that if the choice of a particular type of spin 1 experiment is not a function of the information accessible to the experimenters, then its outcome is equally not a function of the information accessible to the particles. We show that this result is robust, and deduce that neither hidden variable theories nor mechanisms of the GRW type for wave function collapse can be made relativistic. We also establish the consistency of our axioms and discuss the philosophical implications. 
\endabstract

\maketitle

\parskip .20cm

\section{Introduction}
Do we really have free will, or, as a few determined folk maintain, is it
all an illusion?  We don't know, but will prove in this paper that if
indeed there exist any experimenters with a modicum of free will, then
elementary particles must have their own share of this valuable
commodity.

``I saw you put the fish in!" said a simpleton to an angler who had used a minnow 
to catch a bass.  Our reply to an analogous objection would be that we use only a minuscule amount of human free will to deduce free will not only of the particles inside ourselves, but all over the universe.

\renewcommand{\thefootnote}{\fnsymbol{footnote}}

To be more precise, what we shall show is that the particles' response\footnote{More precisely still, the universe's response in the neighborhood of the 
particles.} 
to a certain type of experiment is not determined by the entire previous
history of that part of the universe accessible to them.  The free will we
assume is just that the experimenter can freely choose to make any one of a
small number of observations.  In addition, we make three physical
assumptions in the form of three simple axioms.

The fact that they cannot always predict the results of future experiments
has sometimes been described just as a defect of theories extending quantum 
mechanics. However, if our physical axioms are even approximately true, the free 
will assumption implies the stronger result, that {\em no} theory, whether it extends quantum mechanics or not, can correctly predict the results of future spin experiments.  It also makes it clear that this failure to predict is a merit rather than a defect, 
since these results involve free decisions that the universe has not yet made.

Our result is by no means the first in this direction.  It makes use of
the notorious quantum mechanical entanglement brought to light by
Einstein, Podolsky, and Rosen, which has also been used in various forms
by J. S. Bell, Kochen and Specker, and others to produce no-go theorems
that dispose of the most plausible hidden variable theories. Our theorem seems 
to be the strongest and most precise result of this type, and in particular implies
that there can be no relativistically invariant mechanism of the GRW-type (see Section 10) that explains the collapse of the wave function.  

Physicists who feel that they already knew our main result are cautioned that it cannot be proved by arguments involving symbols such as $<,|,>,\Psi,\otimes$, since these presuppose a large and indefinite amount of physical theory.

\subsection {Stating the theorem.}

We proceed at once to describe our axioms.

There exist ``particles of total spin 1'' upon which one can
perform an operation called ``measuring the square of the
component of spin in a direction  $w$'' which always yields one of the
answers 0 or 1.$^1$\footnote{Superscript numbers refer to the corresponding Endnotes.}

 We shall write $w \to i$ ($i = 0$ or $1$) 
to indicate the result of this operation.
We call such measurements for three mutually orthogonal directions $x,y,z$
a {\it triple experiment for the frame} $(x,y,z)$.

\noindent {\bf The SPIN axiom:}  A triple experiment for the frame $(x,y,z)$ always yields the
outcomes $1, 0, 1$ in some order.

We can write this as:  $x \to j$, $y \to k$, $z \to \ell$,
where $j, k, \ell$ are $0$ or $1$ and $j+k+\ell =2$.

It is possible to produce two distantly separated spin 1 particles
that are ``twinned,'' meaning that they give the same answers to
corresponding ques\-tions$^2$.
A symmetrical form of the TWIN axiom would say that if the same triple
$x,y,z$ were measured for each particle, possibly in different orders,
then the two particles' responses to the experiments in individual
directions would be the same. 
For instance, if measurements in the order $x,y,z$ for one particle
produced
    $x \to 1$, $y \to 0$, $z \to 1$,
then measurements in the order $y,z,x$ for the second particle would
produce
    $y \to 0$, $z \to 1$, $x \to 1$.\footnote{For simplicity, we have 
spoken of measuring $x,y,z$ in that order, but 
nothing in the proof is affected if they are measured simultaneously, as in the
``spin-Hamiltonian" experiment of Endnote 1.}
Although we could use the symmetric form for the proof of the theorem, a
truncated form is all we need, and will make the argument clearer:

\noindent {\bf The TWIN axiom:}.
 For twinned spin 1 particles, if the first experimenter A performs a triple experiment 
for the frame $(x,y,z)$, producing the result 
   $x \to j$, $y\to k$, $z \to l$
while the second experimenter B measures a single spin in direction $w$, then if $w$ 
is one of $x, y, z$, its result is that $w \to j, k, \mbox {or } l$, respectively.

\noindent{\bf The FIN Axiom:}
There is a finite upper bound to the speed with which information can
be effectively transmitted.

This is, of course, a well--known consequence of relativity theory,
the bound being the speed of light.  We shall discuss the notion of
``information'' in Section 3, as also the precise meaning we shall give to
``effectively'' in Section 6. (It applies to any realistic physical transmission.)

FIN is not experimentally verifiable directly, even in principle (unlike SPIN and TWIN$^3$).Its real justification is that it follows from relativity and what we call ``effective causality," that effects cannot precede their causes.

We remark that we have made some tacit idealizations in the above preliminary statements 
of our axioms, and will continue to make them in the initial version of our proof.
For example, we assume that the spin experiments can be performed instantaneously, and
in exact directions.  In later sections, we show how to replace both assumptions and
proofs by more realistic ones that take account of both the approximate nature of
actual experiments and their finite duration.

In our discussion, we shall suppose for simplicity that the finite bound
is the speed of light, and use the usual terminology of past and future light--cones, etc.
To fix our ideas, we shall suppose the experimenter A to be on Earth, while 
experimenter B is on Mars, at least 5 light-minutes away.
We are now ready to state our theorem.

\noindent{\bf The Free Will Theorem} (assuming SPIN, TWIN, and FIN)]. 

{\it If the choice of directions in which to perform spin 1 
experiments is not a function
of the information accessible to the experimenters, then the responses of the
particles are equally not functions of the information accessible to
them.}

Why do we call this result the Free Will theorem?  
  It is usually tacitly assumed that experimenters have sufficient free will to
choose the settings of their apparatus in a way that is not determined by past
history.  We make this assumption explicit precisely because our theorem deduces
from it the more surprising fact that the particles' responses are also not
determined by past history. 

Thus the theorem asserts that if experimenters have a certain property, then 
spin 1 particles have exactly the same property.  Since this property for
experimenters is an instance of what is usually called ``free will," we find it
appropriate to use the same term also for particles.

   We remark that the Free Will assumption, that {\em the experimenters' choice of 
directions is not a function of the information accessible to them}, has allowed us to make our theorem refer to the world itself, rather than merely to some theory of the world.  However, in Section 2.1 we shall also produce a modified version that invalidates certain types of theory without using the free will assumption.

  One way of blocking no-go theorems that hidden variable theories have proposed is
``contextuality''-- that the outcome of an experiment depends upon hidden variables
in the apparatus. For the triple experiment in SPIN, contextuality allows the particle's
spin in the $z$ direction (say) to depend upon the frame $(x,y,z)$. However, since the particle's past history includes all its interactions with the apparatus, the Free Will theorem closes that loophole.
  
\section{The Proof} 

We proceed at once to the proof.  We first dispose of a possible naive
supposition -- namely that ``the squared spin $\theta(w)$ in direction $w$''
already exists prior to its measurement.  If so, the function $\theta$ would
be defined on the unit sphere of directions, and have the property

{(i)} that its values on each orthogonal triple would be $1,0,1$ in some order. 

This easily entails two further properties:

(ii) We cannot have $\theta(x) = \theta(y) = 0$ 
for any two perpendicular directions  $x$ and $y$; 

(iii) for any pair of opposite directions $w$ and $-w$, we have 
 $\theta(w) = \theta(-w)$.  Consequently, $\theta$ is really defined on
 ``$\pm$-directions.''

We call a function on a set of directions that has all three of these
properties a ``101-function."  However, the above naive supposition is 
disproved by {\it the Kochen-Specker paradox for Peres' 33-direction configuration},
namely: 
 \begin{figure}[!h]
\centering
\includegraphics{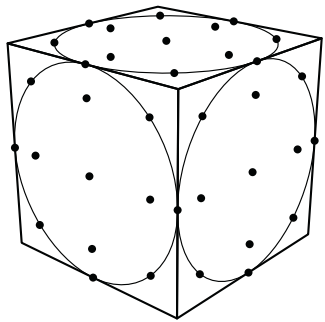}
\caption{\small{The $\pm 33$ directions are defined by the lines joining the center of the cube to the $\pm 6$ mid--points of the edges and the $\pm3$ sets of 9 points of the $3 \times 3$ square arrays shown inscribed in the incircles of its faces.}}
\end{figure}

{\bf Lemma}: 
{\it  There is no 101-function for the $\pm$33 directions of Figure 1.}
  
Since this merely says that a certain geometric combinatorial puzzle has no solution,
for a first reading it may be taken on trust; however we give a short proof in Endnote 4.

\noindent{\bf Deduction of The Free Will Theorem.}

We consider experimenters A and B performing the pair of experiments described 
in the TWIN axiom on separated twinned particles $a$ and $b$, and assert 
that the responses of $a$ and $b$ cannot be functions of all the information 
available to them.

\renewcommand{\thefootnote}{$\star${}}

The contrary {\it functional hypothesis} is that particle $a$'s response is
a function $\theta_a(\alpha)$ of the information $\alpha$ available to it.

In the first instance, we shall suppose that this information is determined by the triple $x,y,z$ together with the information $\alpha'$ that was available just before the choice of that triple, and so is independent of $x,y,z$.  So we can express it as a function
$$
\theta_a (x, y, z; \alpha') =
  \{x \to j, y\to k, z\to \ell \} \, ^\star
  \footnotetext{Here and later we use the fixed symbol $\theta_a$
for this function, despite a change of its variables  (here from $\alpha$ to
$x,y,z ; \alpha'$).}
$$

We refine this notation to pick out any particular one of the
three answers by adjoining a question--mark to the appropriate
one of $x, y, z$; thus:

$$
\aligned
\theta_a(x?, y, z; \alpha' ) & =j \\
\theta_a(x, y?, z; \alpha' ) & =k\\
\theta_a(x, y, z?; \alpha' ) & =\ell\\
\endaligned
$$

Under a similar supposition for we express $b$'s responses as a function
$$
\theta_b (w ; \beta') = \{w \to m \}
$$
of the direction $w$ and the information $\beta'$ available to $b$ before $w$ was
chosen, and again, we write this alternatively as
$$
\hskip -1.25cm \theta_b (w?; \beta')=m.
$$

The TWIN axiom then implies that
$$
\theta_b (w?  ;\beta') =
\left\{
\aligned
  \theta_a (x?, y, z;\alpha')& \qquad \mbox{if } w = x\\
  \theta_a (x, y?, z;\alpha')& \qquad \mbox{if } w = y
\qquad (\star)\\
  \theta_a (x, y, z?;\alpha')& \qquad \mbox{if } w = z\\
\endaligned
\right.
$$

The Free Will assumption now implies that for each direction $w$ and triple
of orthogonal directions $x,y,z$ chosen from our set of $\pm{33}$, there are values of 
$\alpha'$ and $\beta'$ for which every one of the functions in ($\star$) {\it is} defined, since it entails that the experimenters can freely choose an $x,y,z$ 
and $w$ to perform the spin 1 experiments.

Now we defined $\alpha'$ so as to be independent of $x,y,z$, but it is also 
independent of $w$, since there are coordinate frames in which B's experiment happens later than A's.  Similarly, $\beta'$ is independent of $x,y,z$ as well as $w$.
 
Now we fix $\alpha'$ and $\beta'$ and define 
$$\theta_0(w) = \theta_b(w?;\beta'),$$
and find that

$$
\aligned
\theta_a (x?, y, z,\alpha') & = \theta_0 (x) \\
\theta_a (x, y?, z,\alpha') & = \theta_0 (y) \\
\theta_a (x, y, z?,\alpha') & = \theta_0 (z).
\endaligned
$$
 
 Thus $\theta_0$ is a 101-function on the $\pm$33 directions,
in contradiction to the Lemma.  So we have proved the theorem under the
indicated suppositions.

However, one of the particles' responses, say $a$'s, might also depend on some further information bits that become available to it after $x,y,z$ is chosen. If each such 
bit is itself a function of earlier information about the universe (and $x,y,z$)
this actually causes no problem, as we show in the next section.
 
We are left with the case in which some of the information used (by $a$, say) is {\em spontaneous}, that is to say, is itself not determined by any earlier information whatever.  Then there will be a time $t_0$ after $x,y,z$ are chosen with the property that for each time $t < t_0$ no such bit is available, but for every $t > t_0$ some such
bit is available.  

   But in this case the universe has taken a free decision at time $t_0$, because
the information about it after $t_0$ is, by definition, not a function of the
information available before $t_0$!  So if $a$'s response really depends on any
such spontaneous information-bit, it is not a function of the triple $x,y,z$ and the 
state of the universe before the choice of that triple. 

This completes the proof of the Free Will theorem, except for our ascription of the
free decision to the particles rather than to the universe as a whole.  We discuss
this and some other subtleties in later sections after noting the following variant.  

\subsection {\it The Free State Theorem}
As we remarked, there is a modification of the theorem that does not need the Free Will assumption. Physical theories since Descartes have described the evolution of a state from an initial arbitrary or ``free" state according to laws that are themselves independent of space and time. We call such theories with arbitrary initial conditions {\it free state theories}.  

\noindent {\bf The Free State theorem}  ( assuming SPIN, TWIN, FIN.)

{\it No free state theory can exactly predict the results of twinned spin 1 experiments for arbitrary triples $x,y,z$ and vectors $w$.  It fact it cannot even predict the 
outcomes for the finitely many cases used in the proof.}  

This is because our only use of the Free Will assumption was to force the 
functions $\theta_a$ and $\theta_b$ 
to be defined for all of the triples $x,y,z$ and 
vectors $w$ from a certain finite collection and some fixed values $\alpha'$ and
$\beta'$ of other information about the world. Now we can take these as the given 
initial conditions.   
  
   We shall see that it follows from the Free State theorem that no free state theory 
that gives a mechanism for reduction, and {\it a fortiori} no hidden variable theory 
(such as Bohm's) can be made relativistically invariant. 

\section {Information}

Readers may be puzzled by several problems.
In the first place, was it legal to split up information in the way we did in the proof? To justify this, we shall use the standard terminology of information theory,
by identifying the truth value of each property of the universe$^5$ with a 
{\it bit} of information. These truth values are then simply {\it information}, which 
therefore can as usual be thought of as a set of bits.  We emphasize that we do not assume any structure on the set of properties or put any restriction on the simultaneous existence 
of properties.  The only aspect of information that we use is that it consists of set of bits of information, which we can partition in various ways.

Not all information in the universe is accessible to a particle $a$.
In the light of FIN, information that is space--like separated from
$a$ is not accessible to $a$.  The information that {\it is} accessible to $a$ is
the information in the past light cone of $a$.

We redefine $\alpha'$ to be {\em all} the information used by $a$ that is independent
of $x,y,z$, and show that in fact any information-bit used by $a$ is a function of $\alpha'$ and $x,y,z$. For when $x,y,z$ are given, any information-bit $i(x,y,z; {\alpha'})$ that is a function of $x,y,z$ (and maybe some of the earlier information  independent of $x,y,z$ and so in $\alpha'$) is redundant, and can be deleted from the arguments of the function  $\theta_a$. To see this, observe that experimenter A need use only certain orthogonal triples $$(x_1,y_1,z_1), (x_2,y_2,z_2),\ldots ,(x_{40},y_{40},z_{40}),$$
namely the 16 orthogonal triples inside the Peres configuration together with the 24
that are obtained by completing its 24 remaining orthogonal pairs.

Then the information bit $i (x,y,z; \alpha')$ will be one
of the particular bits
$$ i(x_1,y_1, z_1; \alpha'), \ldots , i(x_{40} , y_{40}, z_{40} ; \alpha')
$$
corresponding to these, and since these bits are not functions of the
variables $x,y,z$, they are part of the information $\alpha"$.

Another way to say this is that we are replacing the 
original function $\theta_a$ by a new function 
$$
\theta_a'(x,y,z;\alpha') = 
     \theta_a(x,y,z;\alpha',..., i(x,y,z;\alpha'), ...)
$$
obtained by compounding it with the functions $i$ for each such bit.
  
\subsection {The prompter-actor problem.}

Any precise formulation of our theorem must cope with a certain difficulty 
that we can best describe as follows.  It is the possibility that spin experiments 
performed on twinned particles $a$ and $b$ might always cause certain other 
particles $a'$ and $b'$ to make free decisions\footnote{Our proof dealt with such decisions in the discussion of ``spontaneous information."} of which the responses of $a$ and $b$ are functions. In this context, we may call $a'$ and $b'$ ``promptons," $a$ and $b$ ``actons".  

There is obviously no way to preclude this possibility, which is why we said
that more precisely, it is the universe that makes the free decision in the
neighborhood of the particles. However, we don't usually feel the need for
such pedantry, since the important fact is the existence of the free decision
and that it is made near $a$ and $b$. Let us remind the reader that even the
spin 1 particles $a$ and $b$ are already theoretical constructs, and there is
no point in further multiplication of theoretical entities.  We are really 
talking of spots on a screen, rather than \textit{any} kind of particle$^3$.   

\section{The Consistency Problem for Spin Experiments}

  It cannot be denied that our axioms in combination have some paradoxical aspects.  
One might say that they violate common sense, because $a$ and $b$ must give the same
answers to the same questions even though these answers are not defined ahead of time.  But does that mean that the axioms are logically inconsistent?  This is by no means a trivial question. Indeed, quantum mechanics and general relativity have been mutually inconsistent for most of their joint lifetime, an inconsistency that heterotic string theory resolved (with great difficulty) only by changing the dimension of space-time!
  
   Even the consistency of quantum mechanics with special relativity is somewhat
problematic. Indeed many people (see, e.g., Maudlin[M]) have concluded that when the reduction of the state vector as given by von Neumann's ``Projection Rule" 
is added, paradoxes of the EPR kind contradict relativistic invariance.  So might
our axioms actually be inconsistent?    No!  We can show this using what we shall call a
``Janus model,'' a notion that will at the same time help elucidate some puzzling phenomena. Before we do that, we illustrate the idea by giving a Janus model for an artificially simple construction we call ``hexagonal physics.''

\subsection{A hexagonal universe}

The space--time of this physics is a hexagonal tessellation of the
plane, with time increasing vertically.  An experimenter who is in a
given hexagon on day $t$ can only be in one of the two hexagons that abut it
from above on day $t+1$, the choice between these two hexagons being
left to the experimenter's free will.

\begin{figure}[!h]
\centering
\includegraphics{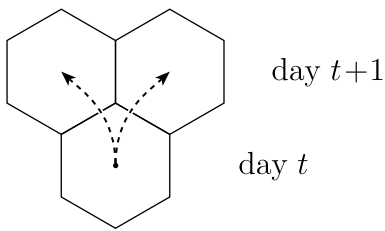}
\caption{Free will in a hexagonal universe}
\end{figure}

We suppose that each hexagon has a ``spin'' whose value $0$ or $1$
can be determined by an experimenter upon reaching that hexagon at a
given day, but not before.  This is the analogue of the FIN axiom.

The only other physical law is that the sum of the spins of three hexagons 
arranged as in Fig. 2 is even (i.e. $0$ or $2$), which is the analogue of 
SPIN (and, as we shall see, also of TWIN, since it relates the spins of
remote hexagons on the same day).  

Are these axioms consistent with each other and with the experimenters'
limited amount of free will?  We can show that the answer is ``yes"
by introducing an agent, Janus, who will realize them.  His realization
will also show that the response of the particles is not a function of past
history in this little universe, showing that they also exhibit a limited
amount of free will according to our definition.

Let us imagine for instance, that two physicists, A and B,
both start at the lowest hexagon of Figure 3 on day $0$,
and  that they never happen to perform their experiments at the same instant. 
 
\begin{figure}[!h]
\centering
\includegraphics{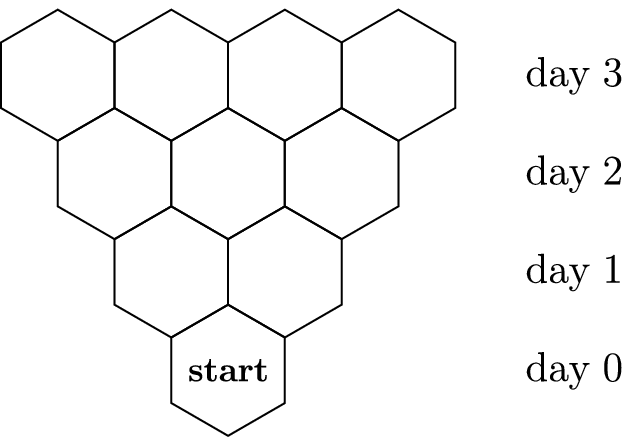}
\caption{}
\end{figure}

  Janus freely decides the result of the first experiment on any given day,
and then uses the SPIN axiom to fill in the results for the other hexagons on
that day.  For example, if on day 5, A and B are at the far left and right hexagons 
of Figure 4 respectively, and the outcome for A on day 5 is 1, then Janus fills in the
other hexagons for day 5 uniquely as in Figure 4 to fulfill the SPIN axiom.

\begin{figure}[!h]
\centering
\includegraphics{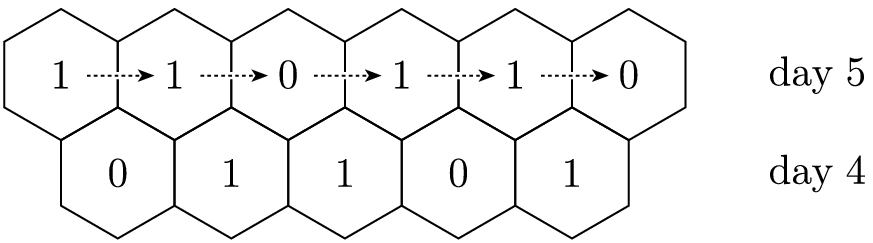}
\caption{}
\end{figure}

The fact that Janus decides on the outcome only at the time of the first
experiment on a given day shows that indeed neither experimenter can
predict the result of an experiment before that day.  SPIN is
also obeyed since Janus uses it to fill in the rest of the hexagons for
that day.

Note that in his realization of hexagonal physics, the speed with which
Janus transmits information is not restricted by our analogue of FIN. Although this 
may seem peculiar, it does not contradict the fact that FIN holds in the model.  It is analogous to the standard way of establishing the consistency of non-Euclidean geometry by constructing a model for hyperbolic geometry (which denies the parallel axiom) inside Euclidean geometry (for which that axiom is true).  The authors have also been greatly influenced by Mostowski's analogous use of the Axiom of Choice to construct a model for 
set theory in which that axiom does not hold. 

Also, Janus need not respect the visible left--right symmetry of hexagonal physics.  
Suppose, for instance that A always moves left, B always moves right,
and that they agree to perform their experiments exactly at noon on each
day.  Then Janus might either use his ``left face'' by freely deciding
the outcome for A and using SPIN to compute the outcome for B,
or use his ``right face" to do the reverse.

If one reader of this paper were to mimic Janus by freely choosing (or
throwing a coin) to determine the spin of either all the leftmost
hexagons in Fig. 3 or all the rightmost ones, and then use SPIN
to fill in the rest, then subsequent readers would not be able to
decide which choice the first reader made.  We can say that this kind of
physics has left--right symmetry even though none of the Janus
constructions do.  Thus, the Janus models show the consistency of this
physics, but cannot be ``the'' explanation for the physics, since there
is both a left and a right Janus model.

Pierre Curie [C] seems to have been the first to enunciate the principle that 
scientific theories should ideally have all the symmmetries of the facts they explain.  Since hexagonal physics has the left-right reflection that its Janus models do not share,
they violate Curie's principle.  In our view, models that violate Curie's principle 
are discredited as explanations, but they do have a proper use, which is to provide consistency proofs.  

Logicians are accustomed to the fact that assertions inside a model often differ
from those outside it.  For example, the ``straight lines" in Poincare's model for
hyperbolic geometry are actually circular arcs, while the ``sets without choice
functions" inside Mostowski's model for set theory actually {\em do} have choice
functions outside it.

  In a similar way, since Janus is not himself part of the physics he realises,  
he is not himself subject to its laws.  His very name might already have suggested 
that we need no longer believe in him!

\subsection{ Consistency of our axioms for spin experiments}
There is a similar Janus model that establishes the consistency of the real 
axioms SPIN, TWIN, FIN, together with the Free Will assumption. 
Janus chooses a co--ordinate frame and
decides his response to the twinned spin--1 experiments of A and B in
the order they happen in this frame.  How does he do this?  The answer is
that he uses a truly random coin, or his own free will (!) to produce
the outcome $0$ or $1$, unless this value is already forced by SPIN and
TWIN, (i.e. $z \to 0$ forces $y \to 1$, $x \to 1$, while $y\to 1$ forces
$z \to 0$, and $x \to j$ for either experimenter forces $x\to j$ for the other.)
Clearly, it is always possible to obey SPIN and TWIN, and the Free Will
assumption holds since neither the decisions of the experimenters nor Janus's 
answers are determined ahead of time.

The possible responses produced by this method are Lorentz invariant,
despite the fact that Janus's method manifestly is not.  The image of
Janus's method under a Lorentz transformation is of course the analogous
method for the image coordinate frame.  Since Janus's method is causal,
this shows that the phenomena appear to be causal from every coordinate
frame. The technical language of Section 6 describes this by 
saying they are ``effectively causal.''  It is 
obvious that the inhabitants of a given Janus model cannot transmit
information backward in time, so by symmetry they cannot effectively
transmit information superluminally --- in other words, FIN holds in
the Janus model. (See the discussion of effective notions in Section 6).

\section{The Consistency of Free Will with quantum mechanics}
In 1952, David Bohm produced a well--known model for quantum 
mechanics (including von Neumann's Projection Rule). 
 This is contentious because Bohm's construction 
(as in fact  he was well aware) does not share the relativistic 
invariance of the physics it ``explains.'' This means that in 
our language it must only be what we have  called a Janus model,
rather than ``the" real explanation of the behavior of the world,
since its images under Lorentz transformations are different equally 
good explanations.  The Free Will theorem shows in fact that 
this construction cannot be made relativistic.

Nevertheless, Bohm's construction was a great achievement, because it is
a Janus model that establishes the consistency of quantum mechanics, 
including the Projection Rule. In fact we can modify it so as 
to prove below the strong result that these are also consistent with the
free will of particles.

\subsection{ Exorcising determinism}
 The main point of hidden variable theories has perhaps been to restore 
determinacy to physics.  Our Free Will theorem is the latest in 
a line of argument against such theories.  However, the situation 
is not as simple as it seems, since the determinacy of such theories 
can be conjured out of existence by a simple semantic trick. 

For definiteness we shall refer to Bohm's theory, which is the best
known and most fully developed one, although the trick is quite general.
According to Bohm, the evolution of a system is completely determined
by certain real numbers (his ``hidden variables''),  whose initial values 
are not all known to us. 

\renewcommand{\thefootnote}{$\star${}}

What we {\em do} know about these initial values may be 
roughly summed up by saying that they lie in a set 
$S_0$.\footnote{More precisely, they will also have a probability distribution $P_0$, which we temporarily ignore.}
An experiment might conflict with some of the initial values, 
and so enable us to shrink the set $S_0$, say to $S_t$ at 
time $t$. The {\em exorcism trick} is just to regard the whole set $S_t$ of
current possibilities, rather than any supposed particular point 
of $S_0$, as all 
that actually  exists at time $t$.

\renewcommand{\thefootnote}{$\dagger${}}

On this view, as $t$ increases, $S_t$ steadily shrinks, not, as Bohm would say, 
because we have learned more about the position of the initial point, 
but perhaps because the particles have made free choices.\footnote{In the more precise version, the probability distribution $P_0$ on the set $S_0$ will be successively refined to more and more concentrated distributions $P_t$ as the time $t$ increases.}

Bohm's theory so exorcised, has become a non--deterministic theory,
which, however, still gives exactly the same predictions!  In fact, the 
exorcised form of Bohm's theory is consistent with our assertion that particles
have free will.  We need only suppose once again that a Janus uses appropriate 
truly random devices to give the probability distributions $P_t$.  
If he does so, then the responses of the particles in our spin experiments, 
for instance, will not be determined ahead of time, and so they will be 
exhibiting free will, in our sense.

As it stands, Bohm's theory visibly contradicts FIN. But since the 
effects it produces are just those of quantum mechanics, they are 
in fact Lorentz invariant.  The exorcised form of Bohm's theory 
therefore performs the service of proving the consistency of 
quantum mechanics (including the Projection Rule) with FIN and the 
Free Will property of particles.
\section{Relativistic Forms of Concepts}

The usual formulations of causality and transmission of information involve 
the intuitive notions of space and time.  Since our axiom FIN is a consequence
of relativity, we must analyse these ideas so as to put them into 
relativistically invariant forms, which we shall denote by prefixing 
the adjective ``effective.''

(i) {\it Effective causality.} The notion of causality is problematic even in 
classical physics, and has seemed even more so in relativity theory. This is
because a universally accepted property of causality is that effects never
precede their causes, and in relativity theory time order is coordinate-frame
dependent.

A careful analysis, however, shows that the proper relativistic notion of
causality is really no more problematic than the classical one. 
This is because all we have the right to demand is that
{\it the universe should appear causal from every coordinate frame}.
We call this property ``effective causality.''

The Janus models that ``explained'' our twinned
spin experiments are causal, and therefore show that the phenomena are 
compatible with effective causality. 
(The same is true of the spin EPR experiment.)

The situation is admittedly odd, since what is a cause in 
the Janus explanation for one frame becomes an effect in that for another.
However, effective causality has the following nice properties:

(1) No observer can distinguish it from ``real'' causality 
(whatever that means).
   
(2) By definition, it is Lorentz invariant.
   
(3) It is the strongest possible notion of causality that 
{\it is} Lorentz invariant.
   
(4) It is provably compatible with SPIN,TWIN,FIN, and the Free Will assumption. 

(ii) {\it Effective transmission of information.} 
There is a similar problem of
extending the notion of transmission of information to the relativistic case.

Obviously, we cannot invariantly say that 
``information is transmitted from $a$ to $b$"
if $a$ and $b$ are space--like separated, 
since then $b$ is earlier than $a$ in some coordinate frames.
If information is {\it really} transmitted from $a$ to $b$, 
then this will appear to be so in all coordinate frames,
which we shall express by saying that 
{\em information is effectively transmitted from $a$ to $b$.}

Many physicists believe that some kinds of information really are transmitted
instantaneously.  We discuss the fallacious argument that suggests this in the
next section.

 (iii) {\it Effective semi-localization.} 
A similar definition can help us understand
where the ``free will'' decision we have found is exercised. 
We shall say that a phenomenon is ``effectively located in a certain 
(not necessarily connected) region of space--time'' 
just when this appears to be so in every coordinate frame.

Then it is clear that we cannot describe the outcome 00 or 11 to 
one of our twinned spin 1 experiments as 
``having been determined near $a$,'' since in some frames it was
known earlier near $b$.  We can, however, say that choice of
00 or 11 is effectively located in some neighborhood of the pair $a,b$ 
(i.e., a pair of neighborhoods about $a$ and $b$).  
We encapsulate the situation by describing the decision as  ``effectively semi-localized.''  

As we already remarked in the Introduction, our assertion that 
 {\em ``the particles make a free decision"}
is merely a shorthand form of the more precise statement that 
 {\em ``the Universe makes this free decision in the neighborhood of the particles."}

It is only for convenience that we have used the traditional theoretical language of
particles and their spins.  The operational content of our theorem, discussed in
Endnote 3,  is that real macroscopic things such as the locations of certain spots on screens are not functions of the past history of the Universe.  From this point of
view it would be hard to distinguish between the pair of statements italicized above.      

 We summarize our other conclusions:
{\em

(1) What happens is effectively causal.
   
(2) No information is effectively transmitted in either 
direction between $a$ and $b$.  

(3) The outcome is effectively semi-localized at the two sites 
of measurement.}

Our definitions of the ``effective'' notions have the great advantage 
of making these three assertions obviously true.  Although they are 
weaker than one might wish, it is also obvious that they are in fact the
strongest assertions of their type that are relativistically invariant.

   Warning --- ``effective so--and--so,'' although it is relativistically 
invariant, is not the same thing as ``invariant so-and-so.''
It would be inappropriate, for instance, to describe the Janus 
explanations of our twinned spin experiments as ``invariantly causal,'' 
since what is a cause in one frame becomes an effect in another. 
The effective notions are more appropriately described as the invariant 
{\em semblances} of the original ones. ``Effective causality,'' 
although it is indeed a relativistically invariant notion, is not 
``invariant causality'' -- it has merely the {\em appearance} of 
causality from every coordinate frame.

We close this section by emphasizing the strange nature of semi-localization.
We might say that the responses of the particles are only ``semi-free"; in a manner of speaking, each particle has just ``half a mind," because it is yoked to the other.  
However, we continue to call their behavior ``free" in view of the ironic fact that 
it is only this yoking that has allowed us to prove that they have any freedom at all!  

What happens is paradoxical, but the Janus models, even though we don't believe them,
show that it is perfectly possible; and experiments that have actually been performed  confirm it. So we must just learn to accept it, as we accepted the earlier paradoxes of relativity theory.

\section{On Relativistic Solecisms}

Many physicists believe that certain kinds of information 
(``quantum information'' or ``phase information'') 
really are transmitted instantaneously.  Indeed, this might almost
be described as the orthodox view, since it follows from a (careless) 
application of the standard formalism of quantum mechanics.

We shall explain the fallacious argument that leads to this conclusion for the
``spin EPR'' case of a pair A,B of spin 1/2 particles in the singleton 
state $|\uA^\ra_z \rangle |\dA^\rb_z \rangle - |\dA^\ra_z \rangle|\uA^\rb_z \rangle$. 
It says that ``when the measurement of A in direction $z$ yields spin up, 
the state is changed by applying the projection operator 
$P_z \otimes I$ to the singleton state, which
annihilates the second term, so that the state becomes 
$|\uA^\ra_z \rangle |\dA^\rb_z \rangle$ , in which B is spin down.''

The word ``becomes'' in this statement is then misinterpreted to mean 
``changes at the instant of measurement", even though this is, of course, 
relativistically meaningless.
However, all that is really asserted is that if this measurement finds 
A to be spin up, then if and when a similar measurement is also performed on B, B will be found to be spin down.

The assertion that ``B {\em is} spin down" 
(made after A has been found to be spin up)
is grammatically incorrect.  We call it a {\em relativistic solecism}. 
It is important to avoid making such mistakes, 
since they can lead to genuine errors of understanding.
How can we do so?

One easy trick is to use the correct tense for such assertions, which is often
the {\em future perfect} (``will have"). 
A grammatically correct version is that 
if and when {\it both} measurements have been performed, 
they will have found that
A was spin up if and only if they will have found B to be spin down. 
This is a Lorentz invariant way of stating exactly the same facts. 

Figure 5 describes the situation.  An observer C whose past light cone contains
both experiments can legitimately say that 
``A found spin up, B spin down."  However,  A can only say 
that ``if the B measurement has been performed, it
{\em will have} found spin down."  
In this, the ``{\em will}" looks forward from 
A to C, while the ``{\em have}" looks backward from C to B.

\begin{figure}[!h]
\centering
\includegraphics{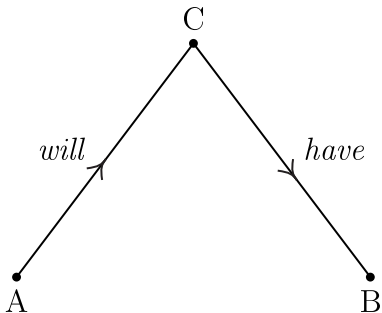}
\caption{A concludes what B {\em will have} found}
\end{figure}

Notice that this makes no mention of the relativistically 
non-existent notion of ``instantaneity," and that (consequently) 
it works equally well for frames in which
the B measurement precedes the A one.  
In fact, it is independent of frame. 
The avoidance of relativistic solecisms is a valuable habit to cultivate!

\subsection{ A Modest proposal}

This line of thought naturally leads us to recommend our 
``Modest Proposal" for the interpretation of states in quantum mechanics. 
According to this, what is usually called the state is merely
a predictor (with probabilities) of what will happen {\em if} various
experiments are performed$^6$. Even when the prediction 
is that some assertion has probability 1, that assertion is 
still contingent on the appropriate experiment's being performed.

Thus if a triple experiment has found  $x \to 1, y \to 1, z \to 0$, 
we certainly know that $S_x^2 = S_y^2 = 1$, but many physicists would 
say that ``we also know $S_w^2 = 1$ for any other direction $w$ 
perpendicular to $z$" (since the probability predicted for this 
assertion is 1).  More modestly, we would say only that ``{\em if 
a measurement is made in direction $w$, it will find $S_w^2 = 1$.}"

To say, in these circumstances, that $S_w^2$ is already $1$, is, in our view,
to be guilty of a simple confusion.  After all, one does not say that 
an astronomical event like an eclipse has already happened as soon as 
it has been predicted with certainty.
   
We revert to the spin EPR case discussed above, supposing that 
a measurement of $\ra$ at time $t$ produces ``spin up", 
giving  $|\uA^\ra_z \rangle |\dA^\rb_z \rangle$ for the state of
the pair, and $|\dA^\rb_z \rangle$ for the state of $\rb$. 
Then we allow ourselves to say that
``$\ra$ {\em is} spin up", since the measurement has actually 
been performed, but not that
``$\rb$ is spin down" at time $t$.  

If the appropriate measurement of $\rb$ is actually performed at time $t$, it
of course produces ``spin down".  But (supposing that $\ra$ and $\rb$ are 
5 light-minutes apart), it {\em will} equally produce ``spin down" 
if it is instead performed 1 minute hence, at time $t + 1$, while if it 
{\em was} performed already at time $t - 1$, it already {\em did}
 produce that answer.  Nothing about $\rb$ changed at time $t$.

Those who would say more might not make any mistaken predictions, but their 
opinions about what happens are not consistent with relativity theory, 
unlike our more modest ones.  As with our discussion of effective notions,
careful speech pays off -- our assertions are obviously both true and
relativistically invariant, while stronger ones are not. 

\section{The Free Will Theorem is Robust}

Our first versions of SPIN and TWIN were tacitly idealized; we now remove some of
this idealization.

In practice, we expect to find deviations from these axioms, 
for instance because the vectors
$x,y,z$ will only be nominally, or approximately, orthogonal, 
rather than exactly so; similarly $w$ will
at best be only only nominally parallel to one of them, 
and again, the twinned pair
might only be nominally in the singlet state. Also, the two theories of
quantum mechanics and special relativity from which we 
derived our axioms, might only be
approximately true. In fact, general relativity is already a more exact theory than special 
relativity. However, we may safely assume:

SPIN$^\prime$: If we observe the squared spin in three nominally 
orthogonal directions, then the probability of a ``canonical outcome" 
(i.e., $j,k,l$ are 1,0,1 in some order) is
at least $1 - \epsilon_s$.

TWIN$^\prime$: If $w$, nominally in the same direction as 
$x$ or $y$ or $z$, yields the value $m$, then the probability that 
$m$ equals the appropriate one of $j,k,l$
is at least $1 - \epsilon_t$.

Then following the argument of the theorem, we define a function $\theta_1(w)$ of direction that behaves like a 101-function in all but a proportion $3\epsilon_t + \epsilon_s$ of cases. For if $w$ is nominally the same as $y$ 
(say), we deduce as before that $$\theta_a(x,y?,z;\alpha') 
= k =_{\epsilon_t}m= \theta_b(w?;\beta'),$$
where ``$=_{\epsilon}$" means ``is equal to except in a 
proportion $\epsilon$ of cases."

Now if we fix on any possible values for $\alpha'$ and $\beta'$ (which exist by
the Free Will assumption) and define $\theta_1(w)$ to be $\theta_b(w?;\beta')$, 
we find 
$$(\theta_1(x),\theta_1(y),\theta_1(z)) 
   =_{3\epsilon_t}(j,k,l) =_{\epsilon_s} 1,0,1$$
in some order.  

But the Lemma shows in fact 
that any function of direction must fail to have the 101--property 
for at least one of 40 particular orthogonal triples 
(the 16 orthogonal triples of the Peres configuration and the
 triples completed from its remaining 24 orthogonal pairs), 
so we have a contradiction unless 
$3\epsilon_t + \epsilon_s \ge 1/40.$

How big may we expect the epsilons to be? Your authors are no experimentalists,
but believe that the errors in angle will dominate the other errors, so that the upper bounds we shall obtain by estimating them conservatively can be relied upon.

If  $x,y,z$ make
angles $\alpha, \beta, \gamma$ with each other in cyclic order, then standard quantum 
mechanical techniques$^7$ give 
$$
(2\cos^2\alpha + 2 \cos^2\beta + 2 \cos^2\gamma - 
   4 \cos\alpha \cos\beta \cos \gamma + \cos^2\alpha \cos^2\gamma)/3,
$$
for the probability of a non--canonical result when we observe 
directions $x,y,z$ in that
order.  If $\alpha,\beta,\gamma$ are all in the interval 
$[\pi/2 - \delta, \pi/2 + \delta]$, this gives 
$$ 
\epsilon_s \le (6\delta^2 + 4\delta^3 + \delta^4)/3.
$$
  
Again, if $w$ makes an angle $\phi$ with one of $x,y,z$, 
then the probability
for the non-canonical result 01 or 10 is $2(\sin^2\phi)/3$, 
so if $\phi$ is in the interval $[-\delta,\delta]$, 
then $\epsilon_t \le 2\delta^2/3$.  Thus
$$
3\epsilon_t + \epsilon_s \le 4\delta^2 + 
(4\delta^3 + \delta^4)/3,$$
which is $\le 1/800$ if $\delta \le$ 1 degree.  
 
This means that the
non-canonical observations  000, 100, 010, 001, 111 for SPIN and 01, 10 for TWIN
can be expected to occur less than once in 800 experiments, 
rather than at least once in every 40 experiments,
as implied by the functional hypothesis. 
A more reasonable bound for $\delta$ might 
be 1 minute, giving the upper bound $1/2900000$ for 
the probability of these non-canonical results.

We remarked above that the change from special  
to general relativity made no difference to our results -- now
is a good time to explain why.  The main difference between 
the two theories is that in a curved space--time one should 
replace ``same direction" by ``directions related by parallel transport" 
in the TWIN axiom.  However, near the solar
system, the curvature of space--time is so small that it was 
extremely hard even to detect,
so that any additional angular errors caused by the 
special relativistic approximation will be utterly negligible 
compared to the 1 degree or 1 minute we have assumed.

The same comment applies to the possible replacement of either 
general relativity or quantum mechanics by some putatively more 
accurate theory, provided this preserves the truth of SPIN$^\prime$ 
and TWIN$^\prime$ for some sufficiently small epsilons.

\section{Historical remarks}

 In the 1960's the Kochen--Specker (K--S) paradox and the Bell Inequality
appeared independently, both showing that certain types of
hidden variable theories are at variance with the predictions of quantum
mechanics.  The K--S paradox showed that the so-called ``non--contextual"
hidden variable theories are impossible, while the Bell Inequality implied 
instead that those that satisfy ``Bell locality'' are impossible.
In the 1970's, Kochen showed via an EPR--type twinning experiment for two
spin 1 particles that in fact Bell locality implied the
non--contextuality condition (see [GHSZ], and Heywood and Redhead [HR] for a discussion).

The advantage of the K--S theorem over the Bell theorem is that it
leads to an outright contradiction between quantum mechanics and the hidden
variable theories for a single spin experiment, whereas the Bell theorem only
produces the wrong probabilities for a series of experiments. The present 
authors have been unable so far to obtain a version of the Free Will theorem 
from Bell's inequalities.

The robustness of the (untwinned) K-S paradox was discussed in two other 
ways by Larsson [L] and Simon, Brukner and Zeilinger [SBZ].    

There have also been improvements on the number of directions needed for
the K--S theorem.  The original version [K--S] used 117 directions.  The 
smallest known at present is the 31-direction set found by Conway and Kochen(see [P]).
Subsequently, Peres [P] found the more symmetric set of 33 that we have used 
here because it allows a simpler proof than our own 31--direction one.

In 1989, Greenberger, Horne and Zeilinger [GHSZ] gave a new version of 
Kochen's 1970's form of the K--S paradox.  They use three spin 1/2 particles in 
place of our two spin 1 ones,  and show that the Bell locality assumption 
leads to an outright contradiction to quantum predictions, without probabilities.

We could prove the Free Will theorem using GHZ's spin 1/2 triplets instead of
our spin 1 twins.  The advantages of doing so are
\labelsep=.2cm
\begin{list}{(i)}{}
\item it shows that spin 1/2 particles are just as much free agents as are 
our spin--1 ones.
\item[(ii)] The argument leading to a contradiction is simpler.
\item[{(iii)}] A version of the experiment has actually been carried out (see [PBDWZ]).
\end{list}

Nevertheless, we have given the twinned spin 1 version for the 
following reasons:
\labelsep=.2cm
\begin{list}{(i)}{}
\item As [GHSZ] note, our twinned spin 1
experiment was suggested by Kochen already in the 1970's.
\item[(ii)]  Conceptually, it is simpler to consider two
systems instead of three.
\item[{(iii)}]  The K--S argument in its present version with 33
directions is now also very simple.
\item[{(iv)}]  An experiment with particles remote enough to verify the Free
Will theorem will probably be realized more easily with pairs than with triples.
\end{list}

The  experiments we described in discussing our theorem are so far only
``gedanken--experiments.''  This is because our Free Will assumption
requires decisions by a human observer, which current physiology tells
us takes a minimum of 1/10 of a second.  During such a time interval
light will travel almost 20,000 miles, so the experiment cannot be done
on Earth.

It is possible to actually do such experiments on Earth if the human
choices are replaced by computer decisions using a pseudo--random
generator, as has already been done for the EPR spin experiment [WJSWZ] 
and suggested for the GHZ experiment by [PBDWZ]. Some other recent experiments
along these lines are described in [SZGS],[GBTZ],[ZBGT].

This delegation of the experimenter's free choice to a computer program,
still leads to a Free Will theorem if we add the 
assumption that the particles are not privy to the details of the
computer program chosen.  Note however that replacing the human choice by a
pseudo--random number generator does not allow us to dispense with the Free 
Will assumption since free will is used in choosing this generator!  The
necessity for the Free Will assumption is evident, since a determined
determinist could maintain that the experimenters were forced to choose
the computer programs they did because these were predetermined at the
dawn of time.

\section{The Theory of Ghirardi, Rimini and Weber}

Ghirardi, Rimini and Weber have proposed a theory [GRW] that attempts to
explain the reduction of the state in quantum mechanics by an underlying
mechanism of stochastic ``hits."  Their theory, as it stands, is visibly not
relativistically invariant, but they hope to find a relativistic version.  
We quote from Bassi and Ghirardi [BG]:

\begin{quote}
\small
``It is appropriate to stress two facts: 
the problem is still an open and a quite
stimulating and difficult one. 
However there seems to be some possibility of carrying 
it on consistently.''
\end{quote}
\normalsize

\renewcommand{\thefootnote}{\fnsymbol{footnote}}

The Free Will theorem shows that this hope cannot be realised\footnote[1]{Nevertheless, relativistic versions of GRW have been claimed; eg. Tumulka[T].}, 
if we reject as fantastic the possibility that the ``hits'' 
that control the particles' behavior also completely
determine the experimenters' actions.

This is because the response of particle $a$, say\footnote{Or perhaps the possible free decison (``prompton") at an earlier 
time $t_0$ that prompted this response---see the proof of the theorem.}, may depend only 
on hits in its past light cone, which
(if they physically exist) have already been incorporated in 
the information $\alpha$ and $\beta$ accessible to it. 
However, our proof of the Free Will theorem shows that the particle's 
response is not a function of this information.

Because the argument is rather subtle, we re-examine the relevant part of the proof
in detail.
  
Let $\alpha_0$ be the information from the hits that influences the behavior 
of particle $a$.  Then by FIN, $\alpha_0$ cannot depend on the direction $w$ since
in some frames this direction is only determined later.  It may depend on $x,y,z$, but as in Section 3 we can write it as a function of $x,y,z,$ and the information $\alpha_0'$
contained in it that is not a function of $x,y,z.$  

Similarly the information $\beta_0$ from the hits that influence particle $b$'s
behavior must already be independent of $x,y,z$, and can be written as a function
of $w$ and the information $\beta_0'$ it contains that is not a function of $w$.
We see that this ``hit" information $\alpha_0'$ and $\beta_0'$ causes no
problems - it is just a part of the information $\alpha'$ and $\beta'$ 
already treated in our proof.
  
Not only does this cover classically correlated information, such as signals from
Alpha Centauri, but it also shows that subtle non-local correlations between the hits 
at $a$ and $b$ cannot help.  We can even let both particles be privy to {\em all} 
the information in $\alpha'$ and $\beta'$.  The only things we cannot do
are to let $a$ be influenced by $w$ or $b$ by $x,y,z$ (so breaking FIN), or to let 
the hits that control the particles' behavior also completely determine the experimenters'
choice of directions, contradicting our Free Will assumption.

\subsection {Randomness can't help}

The problem has been thought to lie in determinism:
\begin{quote}
\small
    ``Taking the risk of being pedantic, we stress once more that from 
     our point of view the interest of Gisin's theorem lies in the fact 
     that it proves that if one wants to consider nonlinear modifications
     of quantum mechanics one is forced to introduce stochasticity and
     thus, in particular, the dynamics must allow the transformations of
     ensembles corresponding to pure cases into statistical mixtures.''
\end{quote}
\normalsize
([BG],p.37)

However, our argument is valid whether the hits are strictly 
determined (the case already covered by Gisin) 
or are somehow intrinsically stochastic. 
In either case, the GRW theory implies that the reduction is 
determined by the hits and so contradicts the Free Will theorem.

To see why, let the stochastic element in a putatively relativistic GRW theory be a sequence of random numbers (not all of which need be used by both particles).  Although these might only be generated as needed, it will plainly make no difference to let them be given in advance.  But then the behavior of the particles\footnote{or of the appropriate ``promptons"} in such a theory would in fact be a function of the information available to them (including this stochastic element) and so its explanation of our twinned spin experiment would necessarily involve superluminal transmission of information between $a$ and $b$.  From a suitable coordinate frame this transmission would be backward in time, contradicting causality.

It is true that particles respond in a stochastic way. 
But this stochasticity of response cannot be explained by putting a stochastic 
element into any reduction mechanism that determines their behavior, 
because this behavior is not in fact determined by any information 
(even stochastic information!) in their past light cones.

\subsection {Summary}

We can summarise the argument by saying first, that the information (whether stochastic or not) that the hits convey to $a$ and $b$ might as well be the same, so long as it is not to break FIN by telling $b$ about $x,y,z$ or $a$ about $w$, and second, that then it might as well have been given in advance. Of course it {\em is} possible to let the particles' behavior be a function of ``promptons," but this merely passes the buck -- even if we call these promptons ``hits," {\em they} must be of a kind that cannot be determined by
previous history, even together with stochastic information.

The same argument shows, again assuming the experimenters' free will, 
that no relativistically invariant theory can provide a mechanism 
for reduction, because that would determine a particle's
behavior, contradicting the fact that it is still free to make its own decision.  Moreover, we have seen that the Free Will assumption is not needed for free state theories:
{\it relativistically invariant theories that purport to provide 
answers at least to all our proposed triple experiments cannot 
also provide a mechanism for reduction.}  

This prevents not only GRW, but any scientific theory of this traditional 
free state type, from providing a relativistically invariant mechanism 
for reduction, even without the Free Will assumption.  The theories that 
purport to do so must deny one of SPIN, TWIN, FIN.
   
We remark that Albert and Vaidman [AV] have made another objection to GRW -- that its explanation of the Stern-Gerlach experiment does not produce sufficiently fast reduction. Bassi and Ghirardi's response [BG] places part of the reduction quite literally in the eye 
of the beholder, which however leads to the concordance problem of the next section (in its
acute form).
 
\renewcommand{\thefootnote}{\fnsymbol{footnote}}

\section{Philosophical Remarks related to the Free Will theorem}

\subsection {On free will}
Let us first discuss the Free Will assumption itself.  What if it is false, and the experimenter is not free to choose the direction in which to orient his apparatus?
We first show by a simple analogy that a universe in which every choice is really Hobson's choice is indeed logically possible. Someone who takes a friend to see a movie he has himself already seen experiences a kind of determinacy that the friend does not. 
Similarly, if what we are experiencing is in fact ``a second showing of the universe movie," it is deterministic even if ``the first showing" was not.

It follows that we cannot {\em prove} our Free Will assumption -- determinism, like solipsism, is logically possible.  Both the non-existence of free agents in determism and the external world in solipsism are rightly conjured up by philosophers as consistent if unbelievable universes to show the limits of what is possible, but we discard them as serious views of {\em our} universe.

It is hard to take science seriously in a universe that in fact controls all the
choices experimenters think they make.  Nature could be in an insidious conspiracy to ``confirm" laws by denying us the freedom to make the tests that would refute them.
Physical induction, the primary tool of science, disappears if we are denied access to random samples.  It is also hard to take seriously the arguments of those who according to their own beliefs are deterministic automata!

We have defined ``free will" to be the opposite of ``determinism" despite the fact that since Hume some philosophers have tried to reconcile the two notions -- a position called {\em compatibilism}.  In our view this position arose only because all the physics known in Hume's day was deterministic, and it has now been outmoded for almost a century by the development of quantum mechanics. However, for the purposes of our paper, we can bypass this hoary discussion, simply by saying that the only kind of free will we are discussing, for both experimenters and particles, is the active kind of free will that can actually affect the future, rather than the compatibilists' passive variety that does not. 

\subsection {Free versus Random?}

Although we find ourselves unable to give an operational definition of either ``free" or ``random," we have managed to distinguish between them in our context, because free
behavior can be twinned, while random behavior cannot (a remark that might also interest 
some philosphers of free will).  Bassi and Ghirardi remark that it follows from Gisin's theorem that their ``hits" must involve a stochastic element in order to make the GRW theory relativistically invariant.  We have shown that what the hits really need is some freedom (to be precise, that they must be at least semi-free). It is for reasons including these that we prefer to describe our particles' behavior as ``free" rather than ``random," ``stochastic," or ``indeterminate." 

\subsection {Interpretation of Quantum Mechanics}

We next describe our own thoughts on the interpretation of Quantum Mechanics,
which have been informed by the Free Will theorem even when not strictly implied
by it.  

We first dismiss the idea, still current in popular accounts although long discounted by most physicists, that a conscious mind is necessary for reduction.  It should suffice to say that there has never been any evidence for this opinion, which arose only from the difficulty of understanding the reduction, but has never helped to solve that problem.
The evidence against it is the obvious {\em Concordance Problem} --- if reduction is in the mind of the observer, how does it come about that the reductions produced by different observers are the same?  This problem is particularly acute for our proposed type of experiment, in which the fact that one observer is on Earth and the other on Mars
causes relativistic difficulties.

Von Neumann's ``Cut Theorem" has sometimes been used to support this belief, since it
shows that any single observer can explain the facts by imagining he performs the reduction, but used in the other direction it actually {\em proves} that there can 
be no evidence for this belief, since the facts are equally explained by supposing
the cut takes place outside him.  The belief is akin to solipsism and has the same drawbacks - it does not respect the
symmetry that the facts are invariant under interchange of observers.

\subsection {Textural Tests}

What, then, causes the reduction to take place?  The Cut Theorem shows that current
quantum mechanics, being linear, cannot itself decide this question. We believe that
the reduction is a real effect that will only be explained by a future physics, 
but that current experiments are already informative.  

Every experimentalist knows that it is in fact extremely difficult to maintain 
coherence --- it requires delicate experiments like those of Mach-Zehnder 
interferometry. Consideration of such experiments has led us to believe that the
criterion that decides between wave-like and corpuscular behavior is what we 
may call the {\em texture} of the surroundings.  Roughly speaking, only sufficiently
``smooth" textures allow it to behave as a wave, while ``rough" ones force it to become
a particle.  

Exactly what this means depends on the circumstances in a way that we do not pretend to understand.  Thus in the interferometric context, the half-silvered beam-splitters 
permit wave-like behavior, so count as smooth, while detectors force the collapse
to a particle, i.e., are rough.  

However, the Free Will Theorem tells us something very important, namely that although a
``rough" texture forces {\em some} decision to be made, it does not actually 
choose {\em which} decision that is.  We may regard such a texture as a tribunal
that may require a particle to answer, but may not force it to make any 
particular answer.  A future theory may reasonably be expected to describe more 
fully exactly which ``textures" will cause reductions, but the Free Will Theorem 
shows that no such theory will correctly predict the results of these reductions:-
 
 {\em Textural tests may demand but not command.}

\subsection {Closing remarks.}

It is our belief that the assumptions underlying the earlier disproofs of 
hidden variables remain problematic.  They involve questionable notions such
as ``elements of reality," counterfactual conditionals, and the resulting
unphysical kinds of locality.  Indeed, in his careful analysis of these 
theories, Redhead[R] produces no fewer than ten different varieties
of locality.
  
One advantage of the Free Will theorem is that by making explicit the necessary Free Will assumption, it replaces all these dubious ideas by a simple consequence, FIN, of relativity.  A greater one is that it applies directly to the real world rather than just to theories.  It is this that prevents the existence of local mechanisms for reduction.

The world it presents us with is a fascinating one, in which fundamental 
particles are continually making their own decisions. 
No theory can predict exactly what these 
particles will do in the future for the very good reason that 
they may not yet have decided what this will be!  Most of their
decisions, of course, will not greatly affect things --- we can 
describe them as mere ineffectual flutterings, which on a large 
scale almost cancel each other out, and so can be ignored.
The authors strongly believe, however, that there is a way our 
brains prevent some of this cancellation, so allowing us to 
integrate what remains and producing our own free will.

The mere existence of free will already has consequences for the philosophy of 
general relativity. That theory has been thought by some to show that ``the flow of time"
is an illusion. We quote only one of many distinguished authors to that
effect: ``The objective world simply is, it does not happen"(Hermann Weyl).
It is remarkable that this common opinion, often referred to as the ``block universe" view, has come about merely as a consequence of the usual way of modeling the mathematics of general relativity as a theory about the curvature of an eternally existing arena of space-time.  In the light of the Free Will theorem this view is mistaken, since the future of the universe is not determined. Theodore Roosevelt's decision to build the Panama Canal shows that free will moves mountains, which implies, by general relativity, that even the curvature of space is not determined. The stage is still being built while the show goes on.

Einstein could not bring himself to believe that ``God plays dice with the world,"
but perhaps we could reconcile him to the idea that``God lets the world run free."

\noindent{\bf Endnotes}. 

The following Endnotes provide more detail about certain technical points.

1.  {\it On measuring squared spins}.  Our assertion that 
$S_x^2, S_y^2, S_z^2$ must take the values $1,0,1$ in some order 
may surprise some physicists, who expect sentences involving 
definite values for $S_x,S_y,S_z$ to be meaningless, 
since these operators do not commute.  However, for a spin 1 
particle their squares do commute.

We can envisage measuring $S_x^2, S_y^2, S_z^2$ by an electrical 
version of the Stern--Gerlach experiment (see  Wrede[W]), 
by interferometry that involves coherent recombination of the beams 
for $S_x = +1$ and $S_x = -1$, or finally by the ``spin-Hamiltonian" 
type of experiment described in [KS], that measures an 
expression of the form $aS_x^2 + bS_y^2 + cS_z^2$. 
An example of a spin 1 system is an atom of orthohelium.

2.  {\it On twinning spin 1 particles}.  To produce a twinned pair of 
spin 1 particles, one forms a pair in ``the singleton state," i.e., 
with total spin $0$.  An explicit description of this state is  
$$
|S_w^a=1 \rangle |S_w^b=-1 \rangle +|S_w^a=-1 \rangle
    |S_w^b=1 \rangle - |S_w^a=0 \rangle |S_w^b=0 \rangle
$$ 
This state is independent of the direction $w$. 
We remark that $S_w^a (= S_w \otimes I)$ and
$S_{w'}^b (= I \otimes S_{w'})$ are commuting operators for 
any directions $w$ and $w'$.

Singleton states have been achieved by Gisin, Brendel, Tittel, Zbinden [GBTZ] 
for two spin 1/2 particles separated by more than 10km. 
Presumably a similar singleton state for distantly
separated spin 1 particles will be attained with sufficient technology. 

3. {\it The operational meanings of various terms.} 
Our uses of the terms 
``spin 1 particle" and ``squared spin in direction $w$" seem to refer to 
certain theoretical concepts.  But we only use them to refer to the locations
of the spots on a screen that are produced by suitable beams in the above
kinds of experiment.

Thus our axioms, despite the fact that they derive from the theories of quantum 
mechanics and relativity, actually only refer to the predicted macroscopic 
results of certain possible experiments.  Our dismissal of hidden variable 
theories is therefore much stronger than the previous ones that presuppose 
quantum mechanics.  From a logical point of view this is very
important, since any use of quantum mechanical terminology necessarily 
makes it unclear exactly what is being assumed.

4. {\it  Proof of the Lemma}.
There is no 101--function for the $\pm 33$ directions of Figure 9.

\begin{figure}[!h]

\includegraphics{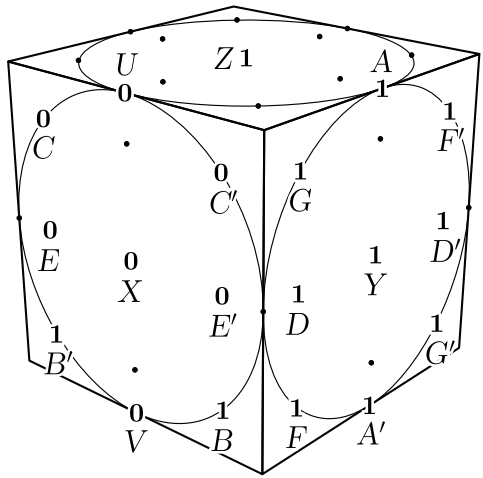}
\caption{Spin assignments for Peres' $33$ directions}
\end{figure}

{\it Proof}.  Assume that a 101--function $\theta$ is defined on these 
$\pm 33$ directions. If $\theta(W) = i$, we write $W \to i$. 
The orthogonalities of the triples and pairs used below in the proof 
of a contradiction are easily seen geometrically.
For instance, in Figure 9, B  and C subtend the same angle at the center O of 
the cube as do  U  and  V, and so are orthogonal.  Thus A,B,C form an 
orthogonal triple.  Again, since rotating the cube through a right angle about 
OZ takes D and G to E and C, the plane orthogonal to  D  passes through 
Z,C,E, so that  C,D is an orthogonal pair and  Z,D,E  is an orthogonal triple.  
As usual, we write ``wlog" to mean ``without loss of generality".

\begin{center}
\begin{tabular}{llll}
\multicolumn{1}{c}{The orthogonality}&
      \multicolumn{2}{c}{}&\multicolumn{1}{r}{and similarly}\\
of $X,Y,Z$ &implies&$X \to 0$, $Y\to 1$, $Z \to 1$ wlog& \\
of $X,A$&implies&$A\to 1$&$A^\prime \to 1$\\
of $A,B,C$&implies&$B\to 1$,$C\to 0$ wlog&$B^\prime \to 1$,
     $C^\prime \to 0$\\
of $C,D$&implies&$D \to 1$&$D^\prime \to 1$\\
of $Z,D,E$&implies&$E \to 0$&$E^\prime \to 0$\\
of $E, F$ and $E,G$&implies&$F \to 1$, $G\to 1$&$F^\prime \to 1$,
     $G^\prime \to 1$\\
of $F,F^\prime, U$&implies&$U \to 0$&\\
of $G,G^\prime, V$&implies&$V \to 0$&
\end{tabular}
\end{center}       
and since $U$ is orthogonal to $V$, this is a contradiction
that proves the Lemma.

5.  {\em On Properties.}
We shall describe the state of the universe or any system in it by
means of properties.  The more usual description in terms of values of
physical quantities such as energy, angular momentum, etc. can always be
reduced to a set of properties, such as ``the energy $E$ lies in the interval 
$(E_1 , E_2)$."  We prefer the more primitive notion of property, because it 
avoids the possible problematic use of the continuum of real numbers in favor 
of $1$ and $0$ (or yes and no), which is more likely to correspond to ultimate 
facts about the world.  More importantly, we have in mind allowing properties
that are more general than allowed by values of physical quantities.

Which properties do we allow?  In classical particle physics the set of
properties is often identified with a Boolean algebra of (Borel)
subsets of a phase space, whereas in quantum mechanics this is replaced
by a lattice of projection operators on Hilbert space. Perhaps we should 
also make some such restriction?

No!  Our theorem would be weakened, rather than strengthened, by any such
restriction.  Also, it is important that we make no theoretical assumptions
about properties, because we don't want our theorem to depend on any physical 
theory.  Our theorem will only be a statement about the real world, as distinct
from some theory of the world, if we refuse to limit the allowed properties in 
any way.  So the answer is: we must allow every possible property!

6. {\em On the Modest Interpretation.}
See Conway and Kochen[CK] for a discussion of this construal of states in the context of
an interpretation of quantum mechanics.  Despite the commonly held view among 
physicists that the ray in Hilbert space contains more information than
probabilities of outcomes, a theorem of Gleason(see [CK]) shows that we can uniquely
characterize rays by these probabilities.

7.  {\em Upper bounds for the epsilons.}
Suppose we make a sequence of measurements of properties with corresponding 
projections $P_1,\ldots ,P_n$ on a system in a pure state $\phi$. 
Then the probability that the properties
all hold is 
$$
\langle P_n \cdots P_1\phi, P_n \cdots P_1\phi\rangle  = 
      \langle \phi, P_1 \cdots P_n \cdots P_1\phi \rangle = 
            tr(P_1 \cdots P_n \cdots P_1P_\phi),
$$ 
where $P_\phi$ is the projection onto the ray of $\phi$. 
This becomes $tr(P_1 \cdots P_n \cdots P_1 \rho)$ if the system is 
in a mixed state given by the density operator $\rho$.

In our case, for SPIN$^\prime$ we have $n = 3$ and $\rho = I/3$, 
since we give equal weight to each of the properties
$P_x,P_y,P_z$ that the squared spin is $0$ in the nominal 
directions $x,y,z$. Then the probability of $000$ for 
$P_x,P_y,P_z$ is 
$$
\begin{aligned}
tr(P_xP_yP_zP_yP_x.I/3) &= 
     tr(|x\rangle \langle x||y\rangle \langle y||z\rangle 
          \langle z||y\rangle \langle y||x \rangle  \langle x|)/3\\
        &= \frac{1}{3} \cos^2\alpha \cos^2 \gamma .
\end{aligned}
$$

Similarly, the probability of 010 is 
$$
tr(P_x(I-P_y)P_z(I-P_y)P_x.I/3) = 
    (\cos^2\beta +\cos^2\alpha \cos^2\gamma 
       -2\cos\alpha \cos\beta \cos\gamma)/3 .
$$
The result in the text is obtained as the sum of five such expressions.

Again, for TWIN$^\prime$, we have $n = 2$ and $\rho = I/3$. 
Observations of a spin 1 particle
(or two twinned particles) in two directions $w,w'$ 
at angle $\phi$ give outcomes 10 or 01 with probability
$$
tr(P_w(I-P_{w'})P_w.I/3) + tr((I-P_w)P_{w'}(I-P_w).I/3) = 
    \rm{\frac{2}{3}}\sin^2\phi .
$$

Our thanks to Eileen Olszewski for the typing and to Frank Swenton
for the graphics.


\begin{thebibliography}{99999999}

\bibitem[M]{M}
Maudlin, T., Quantum Non--Locality and Relativity,
Blackwell, Oxford (1994).

\bibitem[C]{C}
Curie, P.,  {\it J. de Phys.} {\bf 3} (1894), 393--415.

\bibitem[BG]{BG}
Bassi, A., Ghirardi, G. C., {\it Phys. Reports} {\bf 379} (5--6), (2003),
257--426.

\bibitem[T]{T}
Tumulka, R., arXiv: quant--ph/040609v1 (14 Jun 2004).

\bibitem[GHSZ]{GHSZ}
Greenberger, D. M., Horne, M. A., Shimony, A., Zeilinger, A.,
{\it Am. J. Phys.} {\bf 58} (12) (1990), 1131--1143.

\bibitem[HR]{HR}
Heywood, P., Redhead, M. L. G., {Found. of Phys.},
{\bf 13} (1983), 481--499.

\bibitem[KS]{KS}
Kochen, S., Specker, E., {\it J. of Math. and Mech.}
{\bf 17} (1967), 59--87.

\bibitem[P]{P}
Peres, A., Quantum Theory: Concepts and Methods, Kluwer (1993).

\bibitem[PBDWZ]{PBDWZ}
Pan, J.--W., Bouwmeester, D. B., Daniell, M., Weinfurter, H.,
Zeilinger, A., {\it Nature Lett.} {\bf 403} (2000), 515--518.

\bibitem[L]{L}
Larsson, J.--A., {\it Europhys. Lett.} {\bf 58} (6) (2002), 799--805.

\bibitem[SBZ]{SBZ}
Simon, C., Brukner, C., Zeilinger, A., arXiv: quant--ph/0006043v2
(28 Mar 2001).

\bibitem[WJSWZ]{WJSWZ}
Weils, G., Jennewein, T., Simon, C., Weinfurter, H.,
Zeilinger, A., {\it Phys. Rev. Lett.} {\bf 81} (23) (1998),
5039--5043.

\bibitem[GBTZ]{GBTZ}
Gisin, N., Brendel, J., Tittle, W., Zbinden, H.,
{\it Phys. Rev. Lett.} {\bf 81} (1998), 3563.

\bibitem[ZBGT]{ZBTG}
Zbinden, H., Brendel, J., Gisin, N., Tittel, W.,
{\it Phys. Rev. A} {\bf 63}, 022111,  (2001), 1-9.

\bibitem[SZGS]{SZGS}
Stefanov, A., Zbinden, H., Gisin, N., Suarez, A.,
{\it Phys. Rev. Lett.} {\bf 88} 120404,  (2002), 1-4.

\bibitem[R]{R}
Redhead, M.,  Incompleteness, Non--Locality, and Realism,
Clarendon, Oxford (1987).

\bibitem[W]{W}
Wrede, E., {\it Zeits. f. Phys.} {\bf 44} (1927), 261--268.

\bibitem[CK]{CK}
Conway, J., Kochen, S.,
Quantum Unspeakables, (R. A. Bertlmann, A. Zeilinger, eds.)
Springer (2002), pp. 257--270.

\bibitem[AV]{AV}
Albert, D. Z., Vaidman, L.,
{\it Phys. Lett. A} {\bf 139, No. 1,2} (1989), 1-4.



\end{thebibliography}
\end{document}